# Getting beneath the surface of opaque media: universal structure of transmission eigenchannels

Matthieu Davy,[1] Zhou Shi,[2] Jongchul Park,[2] Chushun Tian,[3] and Azriel Z. Genack[2*]

[1] Institut d'Electronique et de Télécommunications de Rennes, University of Rennes 1, Rennes 35042, France.

[2] Department of Physics, Queens College of the City University of New York, Flushing, NY 11367, USA.

[3] Institute for Advanced Study, Tsinghua University, Beijing 100084, China.

*Correspondence to: genack@qc.edu.

**Because the desire to explore opaque materials is ordinarily frustrated by multiple scattering of waves, attention has focused on the transmission matrix of the wave field. This matrix gives the fullest account of transmission and conductance and enables the control of the transmitted flux; however, it cannot address the fundamental issue of the spatial profile of eigenchannels of the transmission matrix inside the sample. Here we obtain a universal expression for the average disposition of energy of transmission eigenchannels for diffusive waves in terms of auxiliary localization lengths determined by the corresponding transmission eigenvalues. The spatial profile of each eigenchannel is shown to be a solution of a generalized diffusion equation. These results reveal the rich structure of transmission eigenchannels and enable the control of wave propagation and the energy distribution inside random media.**

The transmission matrix (TM), whose elements are field transmission coefficients through disordered media, provides a powerful basis for a complete understanding of the statistics of transmission[1-10] and of the degree of control that may be exerted over the transmitted wave by manipulating the incident wavefront[11-22]. Control of the transmitted wave has been demonstrated and characterized in recent years in tight focusing[11-15], and in enhanced or suppressed transmission[16-22] of sound[11], elastic waves[22], light[12-14,16,18,20,21], and microwave radiation[15,19]. The striking similarities in the statistics and scaling of transport of classical and electron waves in disordered systems reflect the equivalent descriptions of transmission and conductance in terms of the TM, $t$[1-10].

The electronic conductance in units of the quantum of conductance is equivalent to the optical transmittance which is the sum over all flux transmission coefficients, $T = \sum_{a,b}^{N} |t_{ba}|^2$, where the $t_{ba}$ are elements of the TM between $N$ incident and outgoing channels, $a$ and $b$, respectively[23]. These channels may be the transverse modes of empty waveguides or momentum states of ideal leads connecting to a disordered sample at a given frequency or energy. The eigenchannels or natural channels of transmission are linear combinations of these channels which can be obtained from the singular value decomposition of the TM, $t = \sum_{n=1}^{N} \mathbf{u}_n \sqrt{\tau_n} \mathbf{v}_n^+$.

Here $\mathbf{u}_n$ and $\mathbf{v}_n$ are unit vectors comprising the $n^{\text{th}}$ transmission eigenchannel and $\tau_n$ are the corresponding transmission eigenvalues[23]. The transmittance may be expressed as the sum over the $N$ transmission eigenvalues, $T = \sum_{n=1}^{N} \tau_n$. Moreover, many key statistical properties of transport through random media such as the fluctuations and correlations of conductance and transmission may be described in terms of the statistics of the transmission eigenvalues[2-10].

Notwithstanding the success of the TM in describing the statistics of transmission through disordered systems, it cannot shed light on the disposition of energy *inside* the sample. Recent simulation[24-28] and measurements[22] in single realizations of random samples suggest that the peak of the energy density profile moves towards the center of the sample and the total energy of the corresponding eigenchannel increases as the transmission eigenvalue $\tau$ increases. This is consistent with measurements of the composite phase derivative of transmission eigenchannels which show that the integrated energy inside the sample increases with $\tau$[29]. The possible universality of the structure of each transmission eigenchannels within scattering media remains an unexplored question, which is the subject of this work. In contrast, the average energy distribution over all $N$ eigenchannels can be obtained by solving the diffusion equation yielding a linear falloff of energy inside a diffusive sample governed by Fick's law for particle diffusion. For localized waves, the energy distribution inside the sample is governed by a generalized diffusion equation and deviates dramatically from a linear fall. This deviation can be explained in terms of a position-dependent diffusion coefficient[30-37].

We seek to discover a universal expression for the average over a random ensemble of samples of the spatial distribution of the energy density averaged over the transverse dimensions for eigenchannels with specified transmission eigenvalue $\tau$. This would extend our understanding of universality in wave propagation from the boundaries to the interior of the sample. Aside from its fundamental importance, a universal description of the scaling of energy density in eigenchannel allows for the control of the energy density profile within the sample. This could be exploited to obtain depth profiles of random media in measurements that are sensitive to optical absorption, emission or nonlinearity. The possibility of depositing energy well below the surface would lengthen the residence time of emitted photons in active random systems and thus lower the lasing threshold of amplifying diffusive media below that for traditionally pumped random lasers. The pump threshold for random lasers is traditionally high because the residence times of emitted photons are relatively short as a consequence of the shallow penetration of the pump laser due to multiple scattering[38-40].

We show that the average of the energy density profile within the sample of an individual eigenchannel with transmission $\tau$ is closely related to the generalized diffusion equation[33-40]. The energy density of the perfectly transmitting eigenchannel with $\tau = 1$ is essentially the sum of a spatially uniform background equal to the energy density of the incident and outgoing wave and the product of this background energy density and the probability density for the wave to return to the cross section at given position within the sample. For $\tau < 1$, we find that the energy density can be expressed as a product of the profile of the fully transmitting eigenchannel and a function governed only by the auxiliary localization length, which was previously used to parameterize the corresponding transmission eigenvalue[1,2].

The profiles of the energy density are found from recursive Green's function simulations[41-43], described in Methods section 1*a*. We consider the propagation of a scalar wave in a quasi-one-dimensional (quasi-1D) strip which is locally two-dimensional (2D), with length $L$ much greater than the transverse dimension $L_t$. The precise geometry of the cross section, which is linear in our study, does not influence the results since waves are mixed in the transverse direction within the bulk of the quasi-1D sample[1-10]. The results are expected to remain valid for locally three-dimensional samples. The Green's function $G(r,r')$ is calculated between grid points $r=(0,y)$ and $r'=(x',y')$ on the incident plane $x = 0$ and in the interior of the sample at depth $x'$, with $y$ being the transverse coordinate. The field transmission coefficient at $x'$ between the incoming mode $a$ and outgoing mode $b$ is calculated by projecting the Green's function onto the empty waveguide modes $\phi_n(y)$, $t_{ba}(x') = \sqrt{v_b v_a} \int_0^{L_y} dy \int_0^{L_y} dy' \varphi_b(y) \varphi_a^*(y') G(r,r')$, where $v_a$ is the group velocity of the empty waveguide mode $a$ at the frequency of the wave. The product of $t(x')$ and $\mathbf{v}_n$ yields the field at $x'$ for different channels due to the $n^{th}$ incoming eigenchannel. In particular, summing the square of the field amplitude at $x' = L$ over all $N$ channels gives the transmitted flux $\tau_n$.

The energy density profiles averaged over 500 configurations drawn from a random ensemble with $N=66$ and $L/\xi = 0.05$ for transmission eigenvalues $\tau=1$, 0.5, 0.1 0.001 are shown in linear and semi-log plots in Fig. 1(a,b). The localization length $\xi$ is determined from $<T>=\pi\xi/2L$[8,9], where $<\ldots>$ indicates the average over an ensemble of random samples. For ease of presentation, we normalize the energy density $W_\tau(x)$ so that it is equal to the transmitted flux $\tau$ at the output surface, $W_\tau(L) = \tau$. This is accomplished by dividing the energy density by the average longitudinal component of the wave velocity at the output surface, $v_+$. In general, $W_\tau(x)$ may be associated with the sum of the forward and backward flux divided by $v_+$. The normalized energy density at the incident surface is equal to 2-$\tau$. For the Lambertian probability distribution for waves transmitted through random 2D media without internal reflection at the boundaries, $v_+ = \pi c/4$, where $c$ is the speed of the wave in the medium. A single peak in $W_\tau(x)$ is seen to increase and shift towards the middle of the sample as $\tau$ increases. The average of the energy density profiles over all transmission eigenchannels, $W(x) = N^{-1}\sum_{n=1}^{N} W_{\tau_n}(x)$, is seen in Fig. 1c to fall linearly as expected[25]. Simulations for a shorter and wider sample with $L/\xi=0.03$ are shown in Fig. 1d. The peak values of the energy density are lower in this sample for all values of $\tau$.

We first consider the energy density profile within the sample of the perfectly transmitting eigenchannel. This is seen in Fig. 1 to be of the form,

$$W_{\tau=1}(x) = 1 + F_1(x) \qquad (1)$$

with $F_1(x)$ a symmetric function peaked in the middle of the sample which vanishes at the longitudinal boundaries of the sample. This is reminiscent of the probability density for a wave to return to a cross section at depth $x$ in an open random medium. This probability density in the case where the surface internal reflection vanishes can be obtained from the correlation function, $Y(x, x')$, which is $\iint dy dy' <|G(x, y, x', y')|^2>$ up to an inessential overall factor which depends upon the wave frequency and the average density of states[35], evaluated at $x = x'$. Physically, $Y(x, x')$ is the ensemble average of local energy density averaged over the cross section at $x$ due to a unit plane source located at $x'$. By using exact microscopic methods, it was found[31-35] that $Y(x, x')$ is the solution of the generalized diffusion equation, $-\frac{d}{dx}D(x)\frac{d}{dx}Y(x,x') = \delta(x-x')$ for waves propagating in open random media. We emphasize that, as in conventional diffusion, $D(x)$ is an intrinsic observable independent of the sources and is determined by microscopic parameters of the system such as disorder strength and incident wavelength. The explicit expression of $D(x)$ can be obtained by either exact microscopic theories[31-35] or approximate self-consistent treatments[30,37].

With $D(x)$ given explicitly by either analytical or numerical methods, the generalized diffusion equation leads to a generalization of Fick's law giving the average flux, $-D(x)dW(x)/dx$, and gives $Y(x, x')$ once the appropriate boundary conditions are implemented. We solve the generalized diffusion equation with the boundary conditions, $Y(x=0, x') = Y(x=L, x') = 1/v_+$. Upon setting $x = x'$ in the solution, we find $Y(x, x) = 1/v_+(1+Y_1(x, x))$, with

$$Y_1(x,x)/v_+ = \int_0^x \frac{dx'}{D(x')} \times \int_x^L \frac{dx'}{D(x')} \Big/ \int_0^L \frac{dx'}{D(x')} \qquad (2)$$

being the generalized probability density of return to the cross section at $x$ (see Methods section 2 for derivations).

In the diffusive limit, $D(x)$ reduces to the Boltzmann diffusion coefficient $D_0$ and Eq. (2) yields $Y_1(x, x)/v_+ = x(L-x)/D_0L$. The quadratic form of $Y_1(x, x)$ is the return probability density in the diffusive limit[30-33]. In 2D, the diffusion coefficient is $D_0 = c\ell/2$, where $\ell$ is the transport mean free path. With the energy density rescaled by $v_+$, we obtain $Y_1(x,x) = \pi x(L-x)/2L\ell$. Plots of $F_1(x)$ for three diffusive samples are shown in Fig. 2a. When normalized to their peak values at $L/2$ in Fig. 2b, these profiles collapse to a single curve, $4x(L-x)/L^2$, showing that $F_1(x)$ equals $Y_1(x,x)$ for diffusive waves. The peak value $F_1(L/2)$ is seen in Fig. 2c to increase linearly with $L/\ell$ with a slope found from the expression above for $F_1(x)$, namely $F_1(L/2) = \pi L/8\ell$.

The relationship between $F_1(x)$ and $Y_1(x, x)$ is further explored by comparing these functions for localized waves for which $Y_1(x, x)$ depends upon $L/\xi$. The profiles of $F_1(x/L)/F_1(L/2)$ obtained in simulations for samples with $N=10$ and in 1D samples (Methods section 1b) are seen in Fig. 3a to narrow as $L/\xi$ increases. $Y_1(x,x)$ is computed from Eq. (2) with $D(x)$ determined from the gradient of $W(x)$ found in simulations as shown in Fig. 3b for the same samples as in Fig. 3a. Profiles of $Y_1(x, x)/Y_1(L/2, L/2)$ and $F_1(x)/F_1(L/2)$ are seen in Fig. 3a to match for each of the random ensembles studied, thereby establishing the result, $F_1(x) = Y_1(x,x)$ and $W_{\tau=1}(x) = 1+Y_1(x, x)$. In the Supplementary Materials we show that this result has a counterpart even in single random samples.

We find from simulations for diffusive waves that the energy density profile can be written as a product

$$W_\tau(x) = S_\tau(x)W_{\tau=1}(x), \qquad (3)$$

in which $S_\tau(x)$ is independent of the value of $L/\xi$ and $L/\ell$, as demonstrated in Fig. 4a. In the Supplementary Materials this factorization is proved for both diffusive and localized waves by non-perturbative diagrammatic technique, and the diagrammatic meaning of $S_\tau(x)$ in terms of the interactions caused by dielectric fluctuations within the medium is given. In the case of the perfectly transmitting eigenchannel, $W_{\tau=1}(x) = 1+Y_1(x, x)$ consists of two parts. One is a spatially uniform background energy density. The other, $Y_1(x, x'=x)$, is associated with the probability of return of waves that have reached the cross section at $x$, with the source of uniform strength, $S_{\tau=1}(x) = 1$, provided by the background energy density that enters into the right-hand side of the generalized diffusion equation as the coefficient of the Dirac delta function. For eigenchannels with $\tau < 1$, the strength of the source term $S_\tau(x)$ falls with increasing depth into the sample since waves do not readily penetrate into the bulk. This modifies the generalized diffusion equation to

$$-\frac{d}{dx}D(x)\frac{d}{dx}Y(x,x') = S_\tau(x')\delta(x-x'), \tag{4}$$

and the boundary conditions to $Y(x=0, x') = Y(x=L, x') = S_\tau(x)/v_+$. Solving this equation (see Methods section 2) and setting $x = x'$ in the solution gives $W_\tau(x) = S_\tau(x) + S_\tau(x)Y_1(x, x)$, which is the same as Eq. (3). Thus, we establish the relation between the factorization and the generalized diffusion equation.

We have not found the analytical form for $S_\tau(x)$ for arbitrary $\tau < 1$. Instead, we could conjecture the form by considering the distribution of transmission eigenvalues. Dorokhov[1,2] showed that for quasi-1D scattering media the variation of average conductance with sample length $L$[44] could be understood in terms of the correlated scaling of the transmission eigenvalues in terms of a set of auxiliary localization lengths $\xi_n$ which gives the average transmission eigenvalue of the $n^{th}$ eigenchannel as $\cosh^{-2}(L/\xi_n)$. For $n<N/2$, $\xi_n^{-1} = (2n-1)/(2\xi)$ varies linearly with channel index $n$[8]. For diffusive waves, this corresponds to $\sim<T>$ open eigenchannels with the average transmission eigenvalue larger than $1/e$,[1-3] while for localized waves, this implies that the transmission is dominated by a single eigenchannel and the probability distribution of $L/\xi'$ exhibits peaks at $L/\xi_n$.[45,46] Here the auxiliary localization length $\xi'$ is associated with the transmission eigenvalue $\tau=\cosh^{-2}(L/\xi')$. To find an expression for $S_\tau(x)$, we hypothesize that average intensity profile in each eigenchannel is related to $\xi'$.

For diffusive waves, $S_\tau(x)$ is governed only by $L/\xi'$. A natural assumption is that the analytic continuation of $S_\tau(x)$ inside the sample at the boundaries is $S_\tau(L) = \tau$ and $S_\tau(0) = 2-\tau$ so that $W_\tau(L) = \tau$ and $W_\tau(0) = 2-\tau$. Thus, $S_\tau(x)$ at the boundaries must reduce to $S_\tau(L) = \cosh^{-2}(L/\xi')$ and $S_\tau(0)=2-\cosh^{-2}(L/\xi')$. Perhaps the simplest expression consistent with these conditions is $S_\tau(x) = 2\frac{\cosh^2((L-x)/\xi')}{\cosh^2(L/\xi')} - \tau$. However, this expression does not agree with the results of simulations shown in Fig. 1. Rather, agreement with simulations is only found once we incorporate an empirical function $h(x/L)$ into the expression above, giving

$$S_\tau(x) = 2\frac{\cosh^2(h(x/L)(L-x)/\xi')}{\cosh^2(h(x/L)L/\xi')} - \tau. \tag{5}$$

To find the function $h(x/L)$, we solve Eq. (5) with $S_\tau(x)$ found in simulations for a given value of $\tau$. Surprisingly, we find that $h(x/L)$ obtained in this way for a single value of $\tau$ in one diffusive sample gives good agreement for all $\tau$ in all diffusive samples (Fig. 4a). $h(x/L)$ is plotted in Fig. 4b. $W_\tau(x)$ found from Eqs. (1,3,5) using this form for $h(x/L)$ are plotted as dashed black curves in Figs. 1(a,b,d). These are in excellent agreement with the profiles found in simulations for diffusive waves.

The universal form of energy density profiles of transmission eigenchannels found here apply to all classical and quantum waves in homogeneously disordered samples. Aside from extending the knowledge of the average disposition of energy from the boundary into the bulk, the expression for $W_\tau(x)$ provides a window on eigenchannel dynamics and the density of states. The integral of $W_\tau(x)$ is the delay time for the transmission eigenvalue $\tau$ and gives the contribution of the eigenchannel to the density of states of the sample[29]. The sum of the integrals of $W_\tau(x)$ over all channels is the density of states of the medium. For waves such as light and sound for which it is possible to control the incident waveform, these results make it possible to tailor the distribution of excitation within a random medium. This may allow depth profiling of random media and make it possible to deliver radiation deep into multiply scattered media.

# Methods

## 1. Numerical simulations

### 1a. Recursive Green's function simulations for quasi-1D random systems

We consider a model of the disordered region stripped of inessential nonuniversal elements. The random sample is index matched to its surroundings with index of refraction of unity so that both internal and external reflections are negligible. Fluctuations $\delta\varepsilon(x,y)$ in the position-dependent dielectric constant, $\varepsilon(x,y) = 1 + \delta\varepsilon(x,y)$, are drawn from a rectangular distribution, the width of which determines the strength of disorder and so the average transmittance of the sample. The wave equation $\nabla^2 E(x,y) + k_0^2 \varepsilon(x,y) E(x,y) = 0$ is discretized on a square grid and solved using the recursive Green's function method. The product of the wave number in the leads $k_0$ and the grid spacing is set to unity. Averages are taken over 500 statistically equivalent samples for each ensemble so that precise comparisons can be made with potential forms for energy density profiles. For all values of $\tau$ in all quasi-1D samples, simulations of $W_\tau(x)$ are averaged over a subensemble of eigenchannels with transmission between $0.98\tau$ and $1.02\tau$.

### 1b. Scattering matrix simulations in 1D disordered samples

We carry out scattering matrix simulations in a 1D system to explore the profiles of $W_{\tau=1}(x)$ and $D(x)$. An electromagnetic plane wave is normally incident upon a layered system with fixed number of layers $L$ with alternating indices of refraction between a fixed value $n$ and 1. The layer thickness is drawn from a distribution that is much greater than the wavelength. The mean free path in this system is equal to the localization length, $\ell = \xi$ [47]. $W_{\tau=1}(x)$ is computed from the average of the energy density over a subset of samples with transmission $\tau$ ranging from 0.995 to 1.

## 2. Solving the generalized diffusion equation

Let us multiply both sides of Eq. (4) by $D(x)$. With the change of variable, $x \to z(x) = \int_0^x dx'/D(x')$ which gives $dz=dx/D(x)$, we rewrite Eq. (4) as

$$-\frac{d^2}{dz^2} Y(z,z') = S_\tau(z')\delta(z-z').$$

The boundary condition is correspondingly modified such that $Y(z(0),z') = Y(z(L),z') = S_\tau(z')/v_+$. This is the normal diffusion equation and is readily solved to give

$$Y(z,z') = S_\tau(z') \times \begin{cases} 1/v_+ + (z-z(0))(z(L)-z')/(z(L)-z(0)), & z < z', \\ 1/v_+ + (z'-z(0))(z(L)-z)/(z(L)-z(0)), & z > z'. \end{cases}$$

We note that since $D(x)>0$, the function $z(x)$ is strictly monotonic and is therefore invertible. Taking this into account and substituting the identity, $z(x_2) - z(x_1) = \int_{x_1}^{x_2} dx'/D(x')$, into the expression of $Y(z, z')$, we may rewrite $Y(z, z')$ as

$$Y(x,x') = S_\tau(x') \times \begin{cases} 1/v_+ + \int_0^x \frac{dx}{D(x)} \int_{x'}^L \frac{dx}{D(x)} \Big/ \int_0^L \frac{dx}{D(x)}, & x < x', \\ 1/v_+ + \int_0^{x'} \frac{dx}{D(x)} \int_x^L \frac{dx}{D(x)} \Big/ \int_0^L \frac{dx}{D(x)}, & x > x'. \end{cases}$$

Setting $x=x'$, we find $Y(x,x) = v_+^{-1} S_\tau(x)(1 + Y_1(x,x))$ with the generalized return probability density given by Eq. (2).

**Acknowledgments**

We thank Pier Mello and Boris Shapiro for stimulating discussions. The research was supported by the National Science Foundation (DMR-1207446) and by the Tsinghua University ISRP.

**Author contributions**
A.Z.G. initiated the research and brainstormed with M.D. and Z.S. regarding the forms of the intensity profiles, M.D. and Z. S. carried out the numerical simulations for multichannel samples, M.D. and J. P. carried out the numerical simulations for 1D samples, M.D. and Z.S. analysed the data, C.T. and M.D. suggested the link between $W_\tau(x)$ and the generalized diffusion equation, which was demonstrated by C.T., A.Z.G, M.D., C.T. and Z.S. wrote the manuscript.
J.P. is presently at Chiral Photonics, Inc. 26 Chapin Road, Pine Brook, NJ 07058, USA.

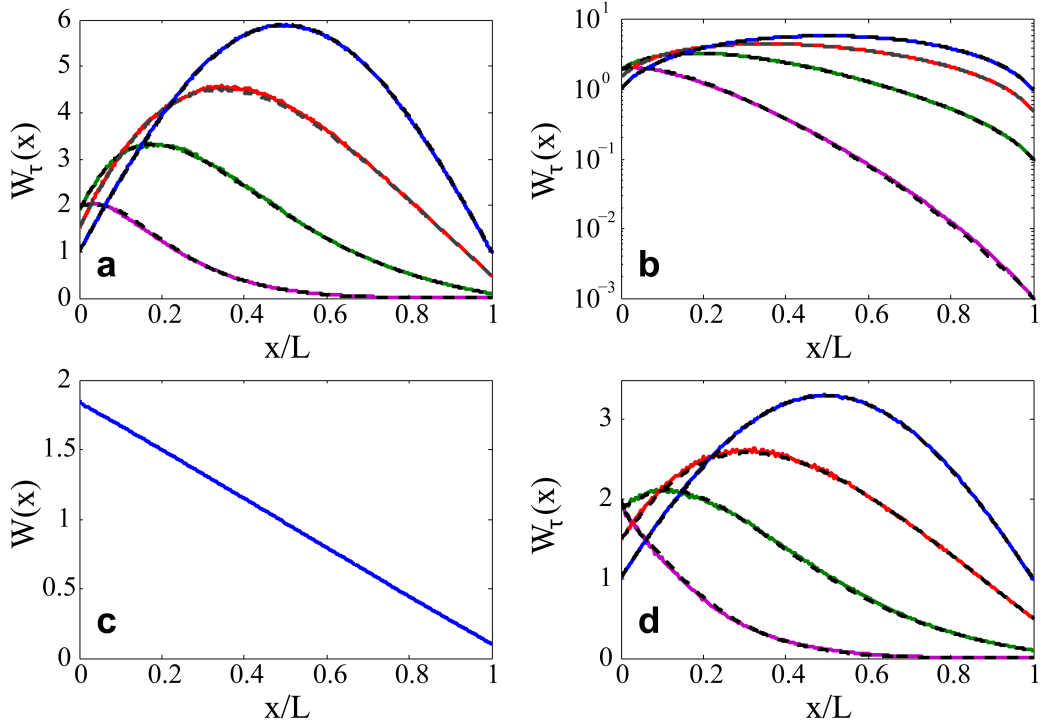

**Figure 1 | Energy density profiles. a**, Ensemble averages of the eigenchannel energy density profiles $W_\tau(x)$ for eigenvalues $\tau=1$, $\tau=0.5$, $\tau=0.1$ and $\tau=0.001$ for a diffusive sample with $L/\xi=0.05$. The black dashed curves are plots of results given by Eqs. (1,3,4). **b**, Semilog plot of data in **a** showing results for small values of the energy density. **c**, Ensemble averages of energy density profiles for all transmission eigenchannels with the eigenchannel indices $n$ from 1 to $N$, $W(x) = N^{-1}\sum_{n=1}^{N} W_{\tau_n}(x)$. The linear falloff of the average of the energy density over all eigenchannels is in accord with the diffusion theory. **d,** $W_\tau(x)$ for a sample with half the width and length as the sample in **a**. Note that, though value of $L/\xi$ is the same as for **a**, the form of $W_\tau(x)$ differs with lower values of the peaks in the shorter sample.

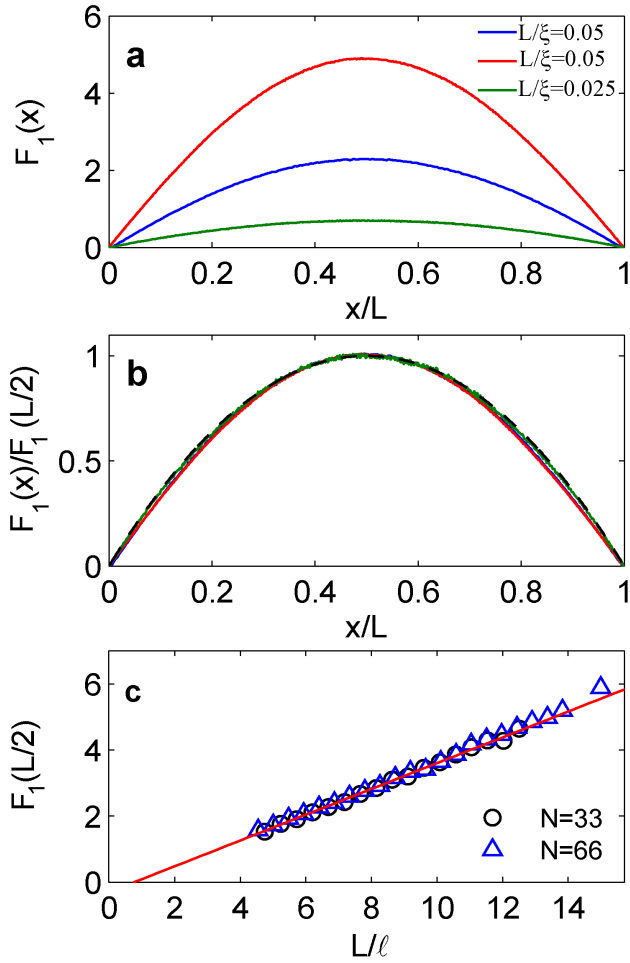

**Figure 2 | Profiles for perfectly transmitting eigenchannels. a,** Ensemble averages of the perfectly transmitting channel $F_1(x)$ for the two diffusive samples of Fig. 1 with $L/\xi$=0.05 and a sample with $L/\xi$=0.025. **b,** $F_1(x)$ normalized by its peak value $F_1(L/2)$. **c,** Scaling of $F_1(L/2)$ with $L/\ell$ for samples with $N$=66 (blue triangles) and $N$=33 (black circles). The linear fit to the results of simulations with a slope of $\pi/8 = 0.39$ given by $Y_1(L/2, L/2)$ gives $F_1(L/2) = (\pi/8)L/\ell - 0.3$ plotted as the red curve.

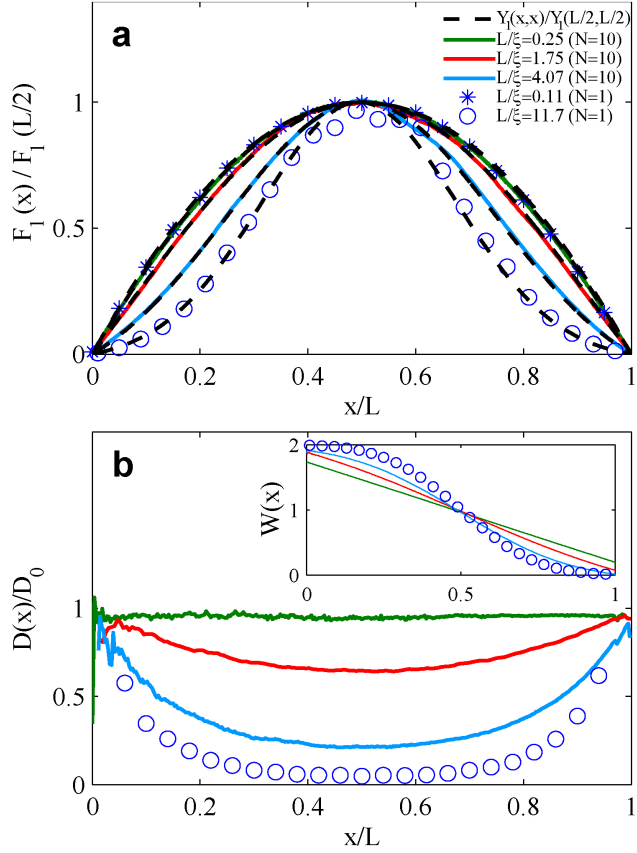

**Figure 3 | Perfectly transmitting eigenchannel profiles and position-dependent diffusion coefficients. a,** $F_1(x)$ normalized by $F_1(L/2)$ for multichannel samples ($N$=10) and for 1D samples ($N$=1) with values of $L/\xi$ ranging from 0.25 to 11.7. The dashed black curves are plots of $Y_1(x,x)$ normalized by $Y_1(L/2,L/2)$ found from Eq. (2) with position-dependent diffusion coefficient $D(x)$ shown in **b**. **b,** Normalized position-dependent diffusion coefficient $D(x)/D(0)$ found from the gradient of $W(x)$ shown in the inset for the same samples as in **a**.

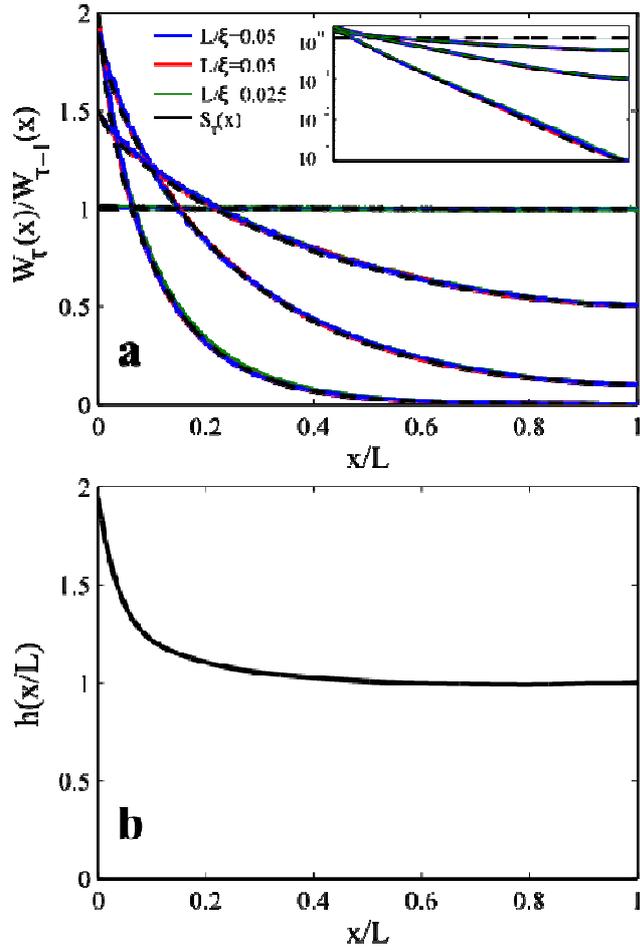

**Figure 4 | Factorization of energy density profiles. a,** $W_\tau(x)/W_{\tau=1}(x)$ for the three ensembles of samples of Fig. 1 (blue and red curves) and a sample with $L/\xi=0.025$ (green curve). The black dashed curves are given by Eq. (5) with $h(x/L)$ shown in **b**. Curves of simulations of $W_\tau(x)/W_{\tau=1}(x)$ for the same values of $\tau$ in different samples and for $S_\tau(x)$ given by Eq. (5) overlap. **b,** $h(x/L)$ determined from Eq. (5).